\documentclass{sig-alternate-05-2015}
\usepackage{graphicx}
\usepackage{subfig}
\usepackage{enumerate}
\usepackage{amsmath}
\usepackage{amssymb}
\usepackage{cases}
\usepackage{amsfonts}
\usepackage{algorithm}
\usepackage{algorithmic}
\usepackage{caption}
\usepackage{url}
\usepackage{multirow}
\usepackage[
  pass,% keep layout unchanged
]{geometry}
\usepackage{environ}
\NewEnviron{myequation}{%
\begin{equation}
\scalebox{0.9}{$\BODY$}
\end{equation}
}

\begin{document}
%
% paper title
% can use linebreaks \\ within to get better formatting as desired
\title{Towards Hybrid Cloud-assisted Crowdsourced Live Streaming: Measurement and Analysis}

%
%\author{\IEEEauthorblockN{Cong Zhang$^{*}$,\ Jiangchuan Liu$^{*}$,\ Haiyang Wang$^{\dag}$}
%\IEEEauthorblockA{$^{*}$School of Computing Science, Simon Fraser University, Canada\\
%$^{\dag}$Department of Computer Science, University of Minnesota Duluth, USA\\
%Email: congz@cs.sfu.ca, jcliu@cs.sfu.ca, haiyang@d.umn.edu}}
\author{
\alignauthor
Cong Zhang$^{*}$, Jiangchuan Liu$^{*}$, Haiyang Wang$^{\dag}$\\
       \affaddr{$^{*}$Computing Science School, Simon Fraser University, BC, Canada}\\
       \affaddr{$^{\dag}$Department of Computer Science, University of Minnesota Duluth, USA}\\
       \email{\{congz, jcliu\}@cs.sfu.ca, haiyang@d.umn.edu}
% 2nd. author
}
\CopyrightYear{2016}
\setcopyright{acmcopyright}
\conferenceinfo{NOSSDAV'16,}{May       13 2016, Klagenfurt, Austria}
\isbn{978-1-4503-4356-5/16/05}\acmPrice{\$15.00}
\doi{http://dx.doi.org/10.1145/2910642.2910644}
% make the title area
\maketitle

\begin{abstract}
%\boldmath
Crowdsourced Live Streaming (CLS), most notably \textit{Twitch.tv}, has seen explosive growth in its popularity in the past few years. In such systems, any user can lively broadcast video content of interest to others, e.g., from a game player to many online viewers. To fulfill the demands from both massive and heterogeneous broadcasters and viewers, expensive server clusters have been deployed to provide video ingesting and transcoding services. Despite the existence of highly popular channels, a significant portion of the channels is indeed unpopular. Yet as our measurement shows, these broadcasters are consuming considerable system resources; in particular, 25\% (\textit{resp.} 30\%) of bandwidth (\textit{resp.} computation) resources are used by the broadcasters who do not have any viewers at all. In this paper, we closely examine the challenge of handling unpopular live-broadcasting channels in CLS systems and present a comprehensive solution for service partitioning on hybrid cloud. The trace-driven evaluation shows that our hybrid cloud-assisted design can smartly assign ingesting and transcoding tasks to the elastic cloud virtual machines, providing flexible system deployment cost-effectively.

%\textbf{C}rowdsourced \textbf{I}nteractive \textbf{L}ive \textbf{S}treaming (CLS) has emerged as a popular content delivery generation allowing every Internet user to live-broadcast content from their personal devices. Such CLS platforms as Twitch.tv provide a series of ingesting and transcoding services bridging a huge number of non-professional broadcasters and massive viewers.
%To offer high-quality CLS service to growing users, the industrial solution is to provision the surplus computation resource and bandwidth capacity from dedicated datacenter. However, our measurement results show that the utilization of dedicated resources is degraded by a large number of unpopular broadcasters, whose workloads are high dynamic and unpredictable.
%
%In this paper, we present a \textbf{Hy}brid \textbf{C}loud-assisted \textbf{C}rowdsourced \textbf{I}nteractive \textbf{L}ive streaming framework HyCIL aiming at alleviating the shortcoming of existing design and optimizing the ingesting and transcoding workloads with the consideration of the characteristics of users' workload during live events. Our work is motivated by the Twitch-based and Amazon EC2-based measurements. After modeling the ingesting and transcoding task assignment problems, we propose two heuristic algorithms to obtain the near-optimal solutions. We evaluate the system performance through trace-based simulations and small-scale test-bed experiments, which demonstrate that our solutions is effective and efficient in terms of deployment cost, streaming quality and broadcast latency.

\end{abstract}
% IEEEtran.cls defaults to using nonbold math in the Abstract.
% This preserves the distinction between vectors and scalars. However,
% if the conference you are submitting to favors bold math in the abstract,
% then you can use LaTeX's standard command \boldmath at the very start
% of the abstract to achieve this. Many IEEE journals/conferences frown on
% math in the abstract anyway.

% no keywords


%%% A category with the (minimum) three required fields
%\category{H.5.1}{Multimedia Information System}{Video}
%%%A category including the fourth, optional field follows...
%\category{C.4}{Performance of Systems}{Measurement techniques}
%\terms{Measurement}
%%
 % NOT required for Proceedings

% For peer review papers, you can put extra information on the cover
% page as needed:
% \ifCLASSOPTIONpeerreview
% \begin{center} \bfseries EDICS Category: 3-BBND \end{center}
% \fi
%
% For peerreview papers, this IEEEtran command inserts a page break and
% creates the second title. It will be ignored for other modes.

%highlight crowdsourced event and broadcast latency
\section{Introduction}

Crowdsourced live streaming (CLS) has emerged as the powerful, real-time means of video broadcasting over the Internet. Such commercial systems as {\it Twitch.tv}\footnote{www.twitch.tv, owned by Amazon.com in September, 2014.} (or Twitch for short) and {\it YouTube Gaming}\footnote{gaming.youtube.com} enable a new form of user-generated live streaming service, attracting an increasing number of viewers all over the globe. Using CLS-based eSports broadcasting as an example, it is known that the number of eSports gaming audiences has trebled to 89 million in the past three years~\cite{report:esports}. The time spent watching eSports broadcasting has increased to 3.7 billion hours in 2014. To fulfill the elevating user demands, CLS service providers are aggressively expending their data center infrastructures. For example, Twitch has already deployed 25 service zones, hosting their dedicated streaming datacenters across five continents.

%Different from peer-to-peer live streaming, where even unpopular live contents still choose stable peers as sources, our Twitch-based measurement demonstrates that the sources of unpopular contents in CLS application also performs more dynamic patterns. Moreover, CLS highlights the event-related live streams with different broadcasters. One representative scenario is that multiple players (i.e., broadcasters) broadcast their game sessions from specific perspectives or languages. In each CLS event, streaming contents have event-based correlation, but show broadcaster-based differences. All these make the workload optimization of CLS challenge.

In this paper, we find that a significant fraction of these expensive data center resources is however consumed by the broadcasters who have very few or even no viewers. In particular, over 25\% (\textit{resp.} 30\%) of the server bandwidth (\textit{resp.} computation) resources are used to host massive broadcasters who do
not have any viewers at all, not to mention other unpopular broadcasters who only have 1 or 2 viewers. To better examine these unpopular broadcasters in CLS systems, we closely monitored 1.5 million broadcasters and 9 million streaming channels within a month. Different from the unpopular channel problem in the traditional streaming systems~\cite{Liu:2009:p2p}, we find that the unpopular broadcasters are not only
in greater numbers but also harder to be managed in CLS systems. In particular, their online behaviors are highly dynamic with short online duration but frequently arrival pattern. These highly dynamic broadcasters (video sources) are not yet considered in the optimization of existing streaming systems. Moreover, CLS highlights the event-related live streams with different broadcasters. One representative scenario is that multiple players (i.e., broadcasters) broadcast their game sessions from specific perspectives or languages. In each CLS event, streaming contents have an event-based correlation, but show broadcaster-based differences. All these make the workload optimization of CLS quite challenging.

Given the dynamic online patterns of these unpopular broadcasters, an intuitive and cost-efficient solution is to migrate their video ingesting and transcoding services to elastic cloud platforms. Our previous EC2-based measurement~\cite{report} also indicates that public cloud can provide comparable transcoding and communication latency if we can smartly assign cloud virtual machines (VMs) to the broadcasters. Based on this observation, we present the design of a hybrid cloud-assisted CLS framework (HyCLS) towards enhancing the utilization of existing dedicated datacenters. We first propose the stable index (SI) to estimate the stability of broadcasters who already have historical activities. We further design effective algorithms to offload the workloads of new broadcasters. The evaluation shows that our proposed solutions can migrate up to 59.9\% of the workloads from dedicated datacenters to public clouds and cost-effectively reduce about 20\% of lease cost in hybrid cloud-assisted design.

\section{Background}
\label{sec:background}
\begin{figure}[!htbp]
  \begin{minipage}[t]{1\linewidth}
  \centering
  \includegraphics[width=0.5\textwidth]{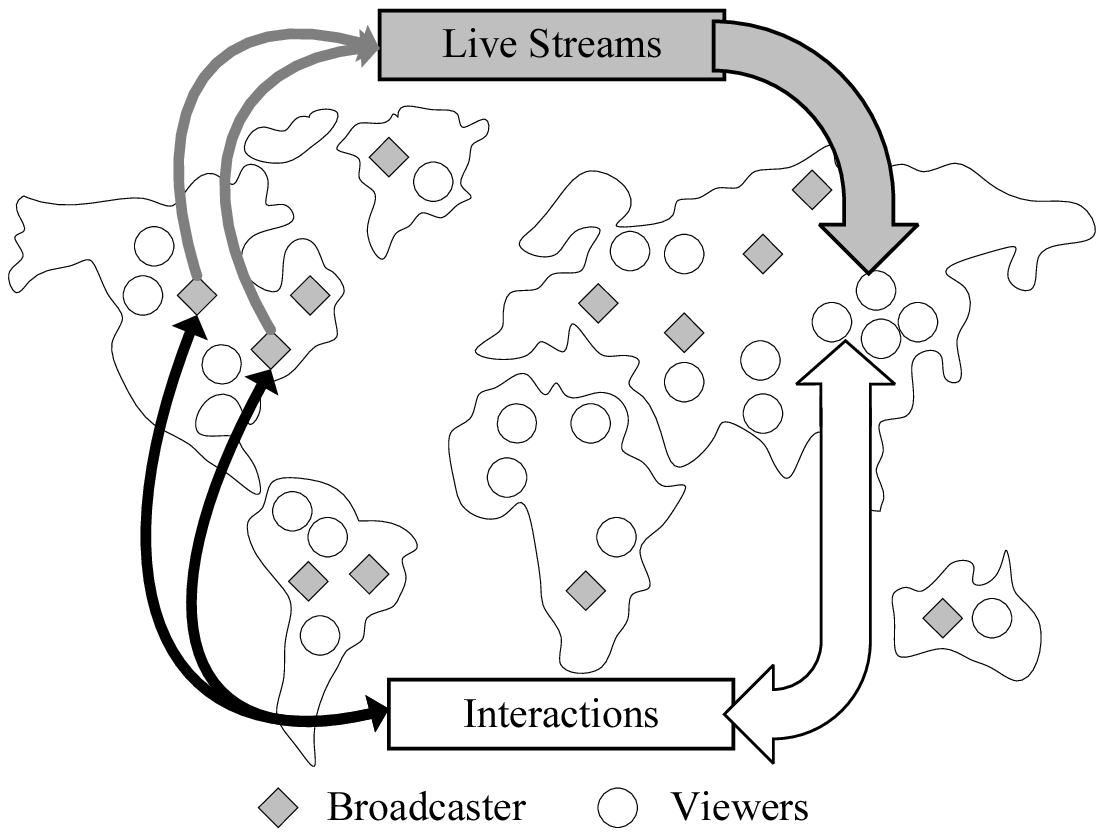}
   \vspace{-3mm}
  \caption{A generic diagram of CLS systems.}
  \vspace{-3mm}
  \label{fig:g}
  \end{minipage}
\end{figure}
The idea of service crowdsourcing has attracted a substantial amount of attentions from both industry and academia~\cite{Wikimapia, Motoyama:2010:USENIX}. This service model refers to the process of getting contributions from a crowd of people (crowdsourcers)~\cite{Motoyama:2010:USENIX}. In multimedia-related crowdsourcing studies, various researchers proposed different frameworks to evaluate and improve users' Quality-of-Experience (QoE) for images and videos processing~\cite{Carlier:2011:MM, Rawat:2014:MM}.

Figure~\ref{fig:g} briefly depicts a generic system diagram of crowdsourced live streaming platforms with the streaming and interactive pipelines that jointly serve geo-distributed broadcasters and viewers. All sources are managed by amateur broadcasters (i.e., crowdsourcers) and driven by massive viewers/broadcasters in real-time using live messages (e.g., TwitchPlaysPokemon). Use Twitch as a case study, our previous work has illustrated the basic architecture of CLS platform~\cite{Zhang:2015:NOSSDAV}. Several recent studies already focused on crowdsourced live streaming services ~\cite{Kaytoue:2012:WWW, Aparicio-Pardo:2015:MMSys}. Different from the existing studies, our work examines the online behaviors of the crowdsourcers and explores the effective utilization of resources. Our work differs from these recent studies in the following aspects: first, we target on crowdsourced live streaming systems which represent lots of unique features. For example, geo-distributed broadcasters determine the service quality from the ``first-mile" of live streaming distribution. The scalability of streams in this process basically determines the viewers' QoE. Second, any improvement and optimization must carefully design the strategy to generate low-latency live streaming. Therefore, we propose an optimal solution that cost-effectively schedules the broadcasters' workloads to public clouds in the crowdsourced live content generation.

\section{Measurement of Crowdsourced Live Streaming: Twitch as a Case Study}
\label{sec:measurementTwitch}
\begin{figure}[!htbp]
\vspace{-3mm}
  \begin{minipage}[t]{1\linewidth}
 \centering
  \subfloat[Daily patterns]{
    \label{fig:event} %% label for first subfigure
    \includegraphics[width=0.4\textwidth]{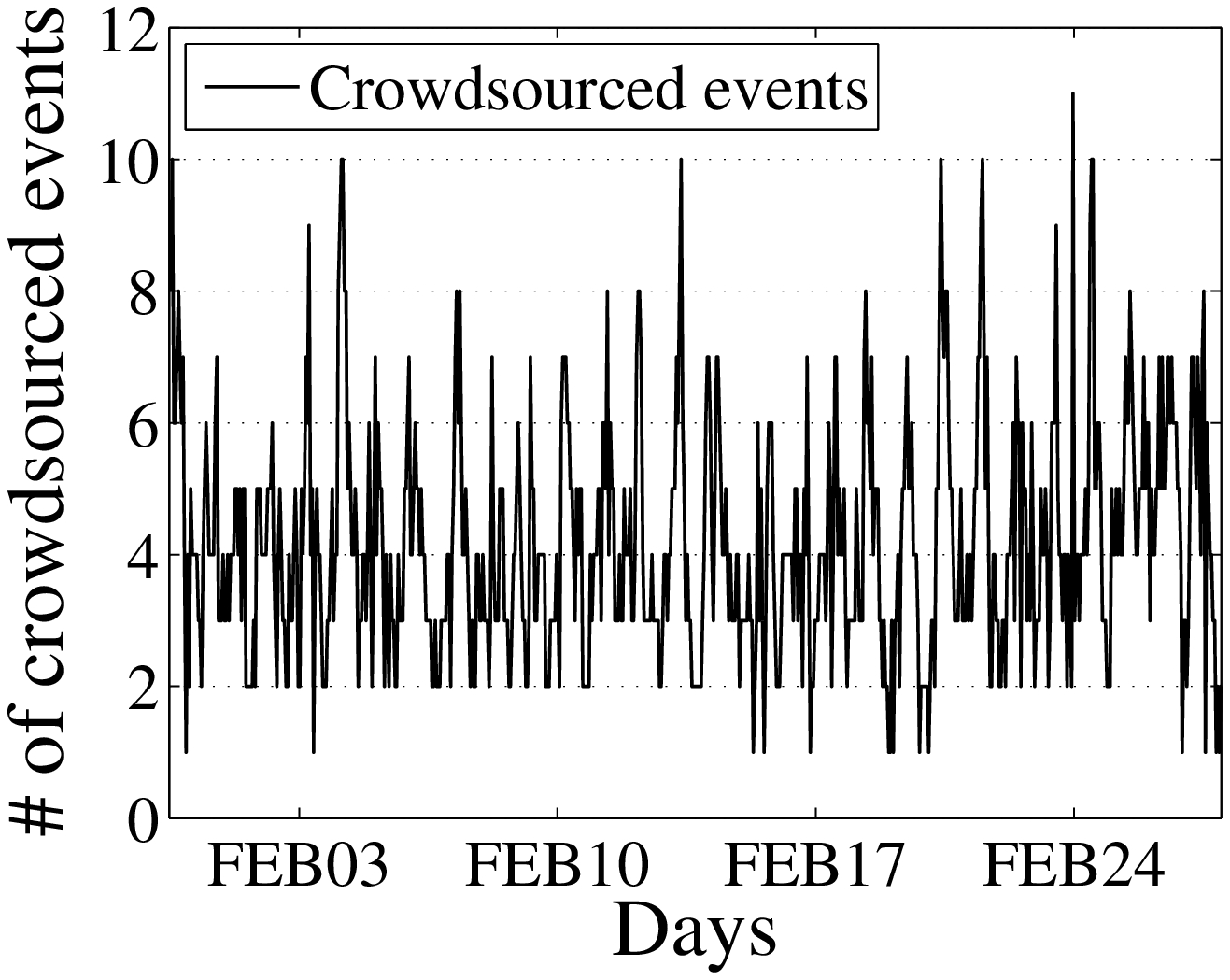}}
  \subfloat[Effects of crowdsourced live events]{
    \label{fig:eventv} %% label for second subfigure
    \includegraphics[width=0.4\textwidth]{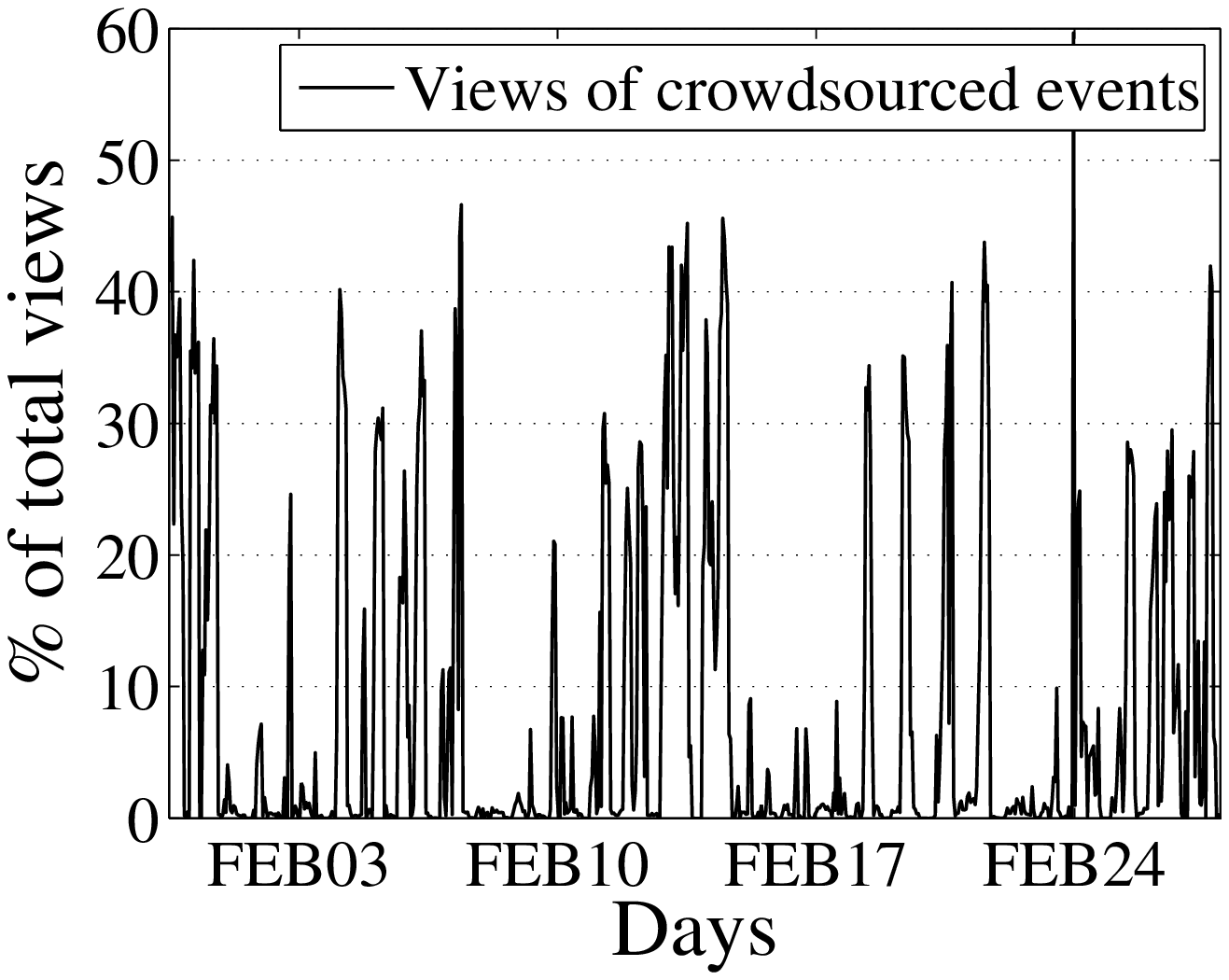}}
    \vspace{-2mm}
  \caption{Characteristics of crowdsourced live events.}
    \vspace{-3mm}
  \label{fig:samples} %% label for entire figure
\end{minipage}
\vspace{-3mm}
\end{figure}

%\begin{figure*}[!ht]
%\vspace{-3mm}
%\begin{minipage}[t]{0.24\linewidth}
%\centering
%\includegraphics[width=1\textwidth]{plotviews.eps}
%\caption{Broadcasters rank ordered by popularity.}
%\label{fig:plotviews}
%\end{minipage}
%\begin{minipage}[t]{0.48\linewidth}
% \centering
%  \subfloat[Popular broadcaster A]{
%    \label{fig:pop} %% label for first subfigure
%    \includegraphics[width=0.5\textwidth]{pop.eps}}
%  \subfloat[Unpopular broadcaster B]{
%    \label{fig:unpop} %% label for second subfigure
%    \includegraphics[width=0.5\textwidth]{unpop.eps}}
%     \vspace{-2mm}
%  \caption{Two samples of broadcasters.}
%  \label{fig:samples} %% label for entire figure
%\end{minipage}
%\begin{minipage}[t]{0.24\linewidth}
%\centering
%\includegraphics[width=1\textwidth]{plotduration.eps}
%\caption{The distribution of duration.}
%\label{fig:plotduration}
%\end{minipage}
%\vspace{-8mm}
%\end{figure*}
\begin{figure*}[!ht]
\begin{minipage}[t]{0.5\linewidth}
 \centering
  \subfloat[Popularity ranking]{
    \label{fig:plotviews} %% label for first subfigure
    \includegraphics[width=0.48\textwidth]{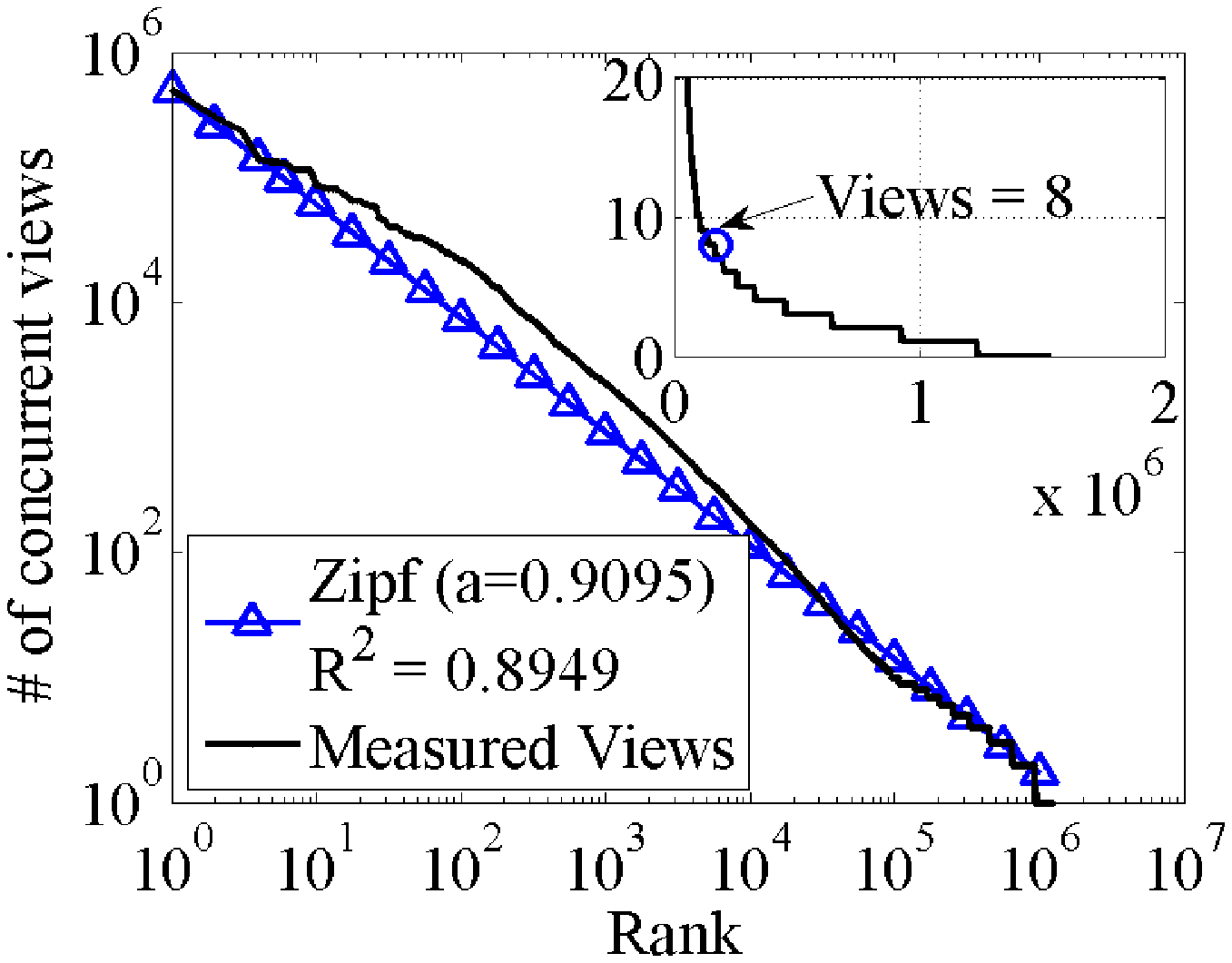}}
  \subfloat[Duration distribution]{
    \label{fig:plotduration} %% label for second subfigure
    \includegraphics[width=0.48\textwidth]{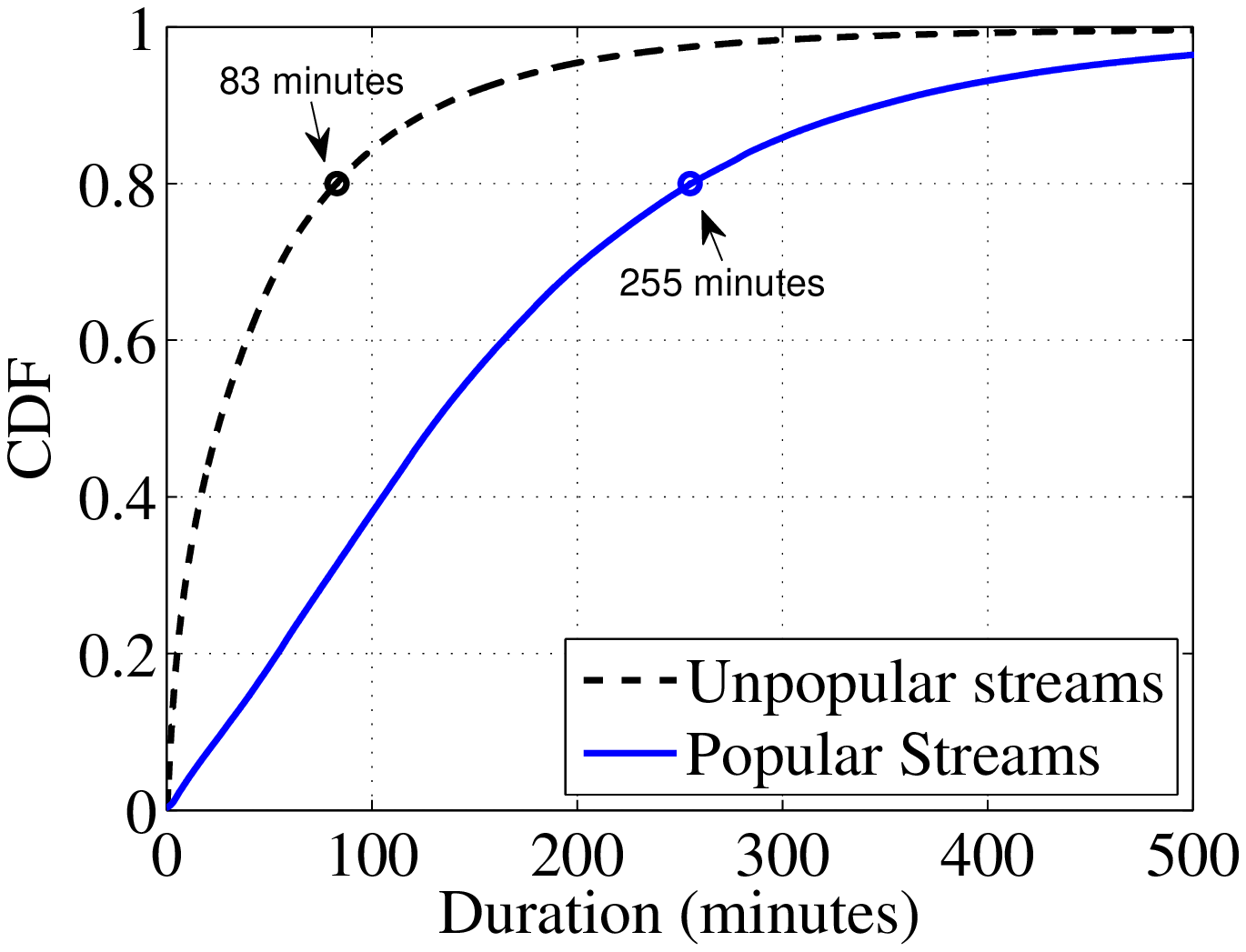}}
     \vspace{-2mm}
  \caption{The characteristics of broadcasters.}
  \label{fig:charbroad} %% label for entire figure
\end{minipage}
\begin{minipage}[t]{0.48\linewidth}
 \centering
  \subfloat[Popular broadcaster A]{
    \label{fig:pop} %% label for first subfigure
    \includegraphics[width=0.5\textwidth]{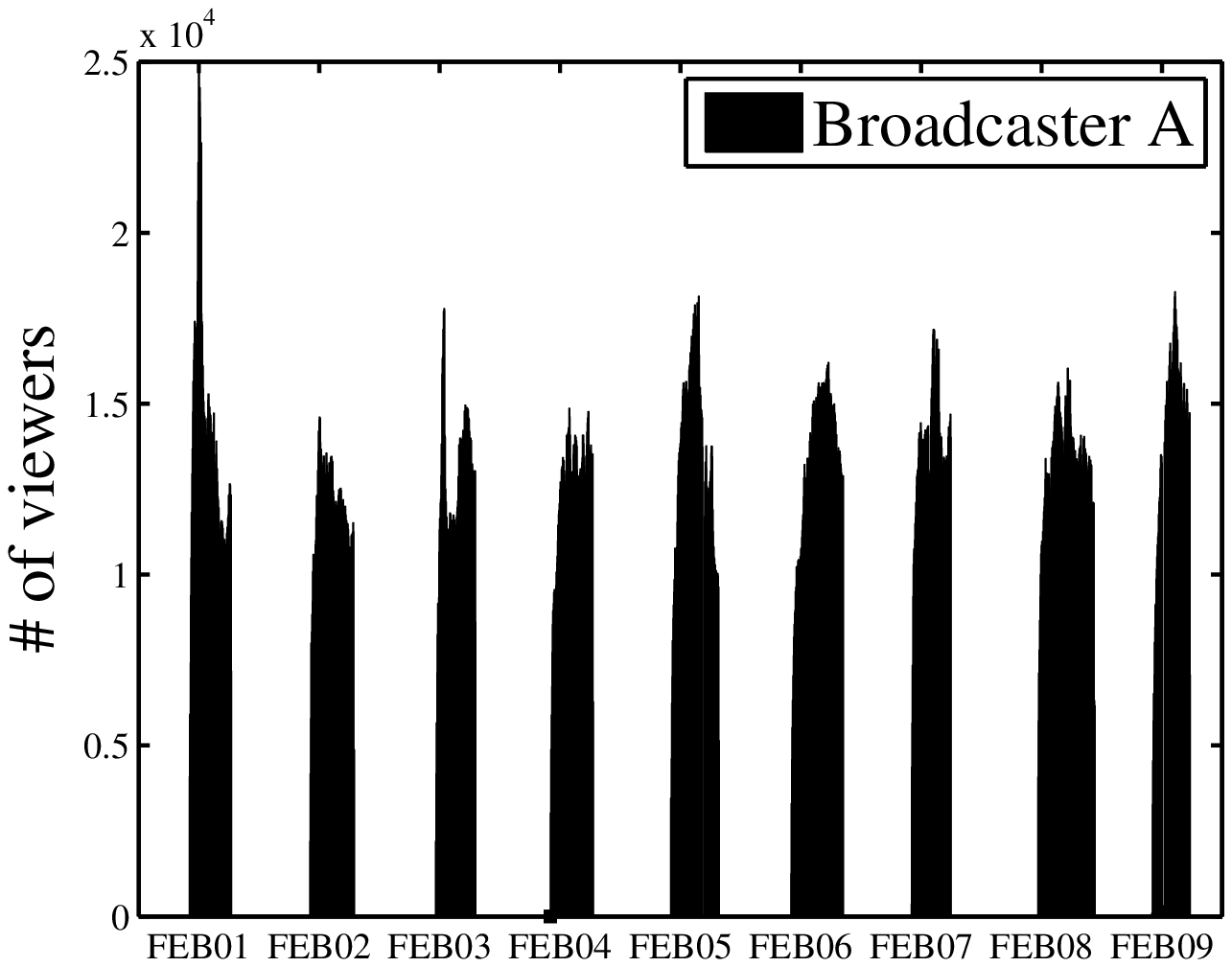}}
  \subfloat[Unpopular broadcaster B]{
    \label{fig:unpop} %% label for second subfigure
    \includegraphics[width=0.5\textwidth]{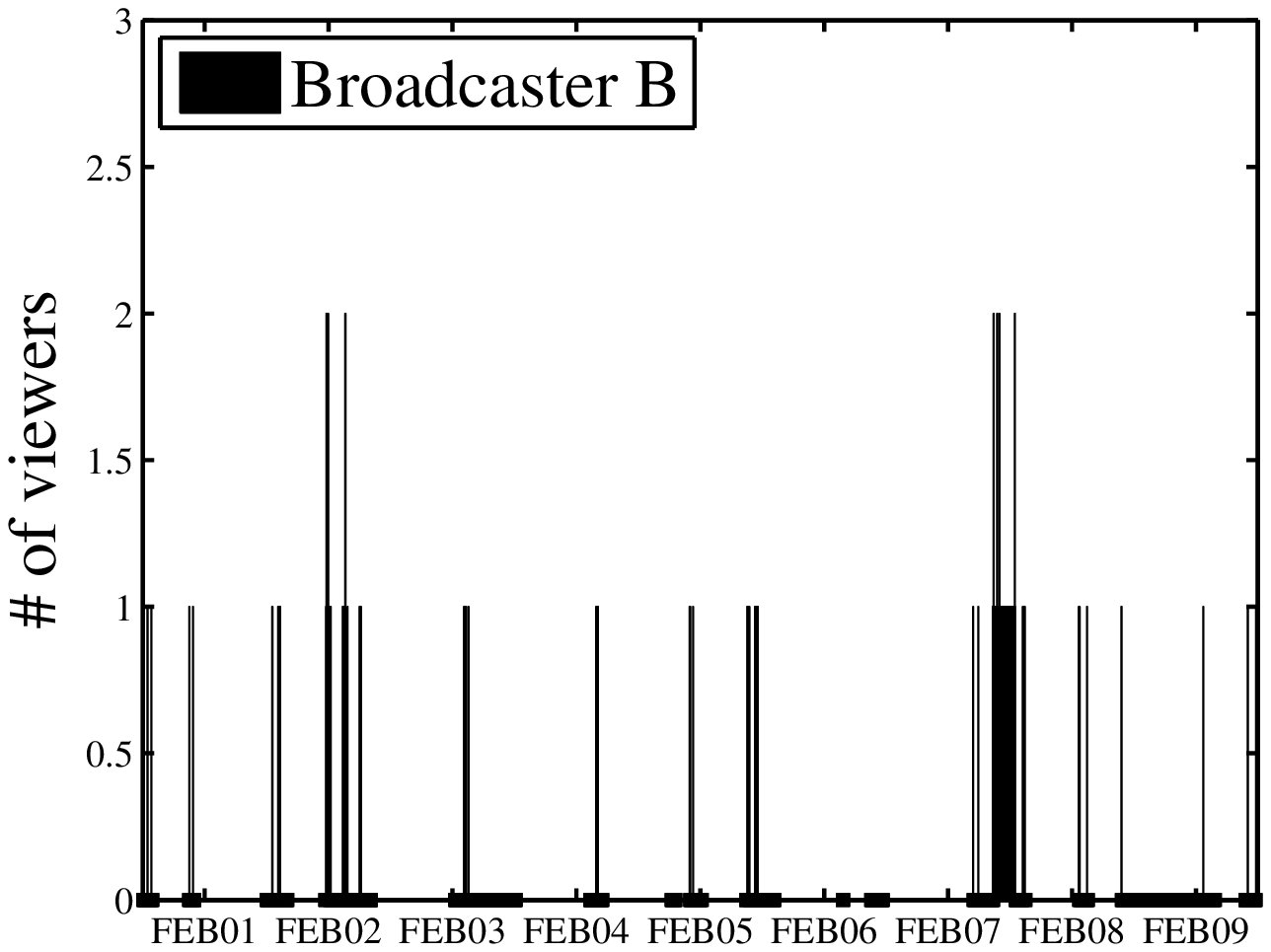}}
     \vspace{-2mm}
  \caption{Two samples of broadcasters.}
  \label{fig:samples} %% label for entire figure
\end{minipage}
\vspace{-8mm}
\end{figure*}
%\begin{minipage}[t]{0.67\linewidth}
%\centering
%\includegraphics[width=0.9\textwidth]{pop.eps}
%\caption{An example of popular broadcaster.}
%\label{fig:pop}
%\end{minipage}
%\begin{minipage}[t]{0.33\linewidth}
%\centering
%\includegraphics[width=0.9\textwidth]{unpop.eps}
%\caption{An example of unpopular broadcaster.}
%\label{fig:unpop}
%\end{minipage}

In our measurements, we try to answer the following fundamental questions: how many unpopular broadcasters exist in real crowdsourced live streaming systems? And, what is the underlying impacts of them? As such, we deeply measure the workloads and the corresponding resource consumption based on the crawled datasets from Twitch.

Our investigation is based on crawled data, which are continually collected from Twitch every five minutes in a one-month period (Feb.1st-28th, 2015). Through the Twitch APIs~\footnote{http://dev.twitch.tv/}, our multi-thread crawler obtained information from each broadcaster and whole system dashboard. We retrieved broadcaster dataset and stream dataset through polishing aforementioned data. A brief explanation is as follows:
\begin{itemize}
\item in broadcaster dataset: each trace collects the total number of views and other statistics such as playback bitrate, resolution, and partner status, for a total of $1.5$ million broadcasters (2\% outliers are eliminated).
%$1,530,578$
\end{itemize}
\begin{itemize}
\item in stream dataset: each trace records the number of viewers every five minutes and other properties including start time, duration, game name, etc., for a total of $9$ million streams (0.3\% outliers are removed).
%$9,048,346$
\end{itemize}

\textit{Effects of Crowdsourced Live Events}: CLS highlights the event-related live streams with different broadcasters. One representative scenario is that multiple players (i.e., broadcasters) broadcast their game sessions from specific perspectives or languages. In each CLS event, streaming contents have an event-based correlation, but show broadcaster-based differences. To illustrate this distinct feature, we explore the crowdsourced live events based on the broadcaster's channel name and game type in each trace of stream dataset. Figure~\ref{fig:event} plots the number of crowdsourced events during one month~\footnote{Due to space limitation, only four date labels are displayed.}. We observe that crowdsourced live events exist in all data traces. Although the highest number is twelve, these live events attract up to 52\% of total views in our measurement, as shown in Figure~\ref{fig:eventv}. This distinct feature will be considered in the problem formulation (Section~\ref{sec:formulation}).

\textit{Popularity of Crowdsourced Streaming Channels}: We then focus on the distribution of broadcaster's popularity, which has played a key role in previous studies for multimedia systems, and is also critical to answer our first question. We plot the highest number of concurrent views against the rank of the broadcasters (in terms of the popularity) in log-log scale in Figure~\ref{fig:plotviews}.
From this figure, we observe that the popularity of those broadcasters exhibits perfect Zipf's pattern\footnote{We use the coefficient of determination, denoted $R^2$, to illustrate how well our measured data fit the Zipf's law.}. We further find that there exists such a high skewness that the top-3\% popular broadcasters account for about 80\% of the peak requests.
Another interesting result shows that 90\% of broadcasters only attract less than 8 viewers (labeled on the small figure in Figure~\ref{fig:plotviews} ) even at their peak time in our one-month broadcaster dataset. Based on these findings, if the highest number of concurrent views number of a broadcaster is less than 8, we assume that s/he is not a popular broadcaster. How long are these broadcasters' live streams? Is there any difference between popular and unpopular broadcasters in terms of live duration? We next investigate the characteristics of broadcasters' streams based on stream dataset.
\begin{figure*}[!ht]
\begin{minipage}[t]{0.5\linewidth}
 \centering
  \subfloat[Popular broadcasters]{
    \label{fig:arrivala} %% label for first subfigure
    \includegraphics[width=0.48\textwidth]{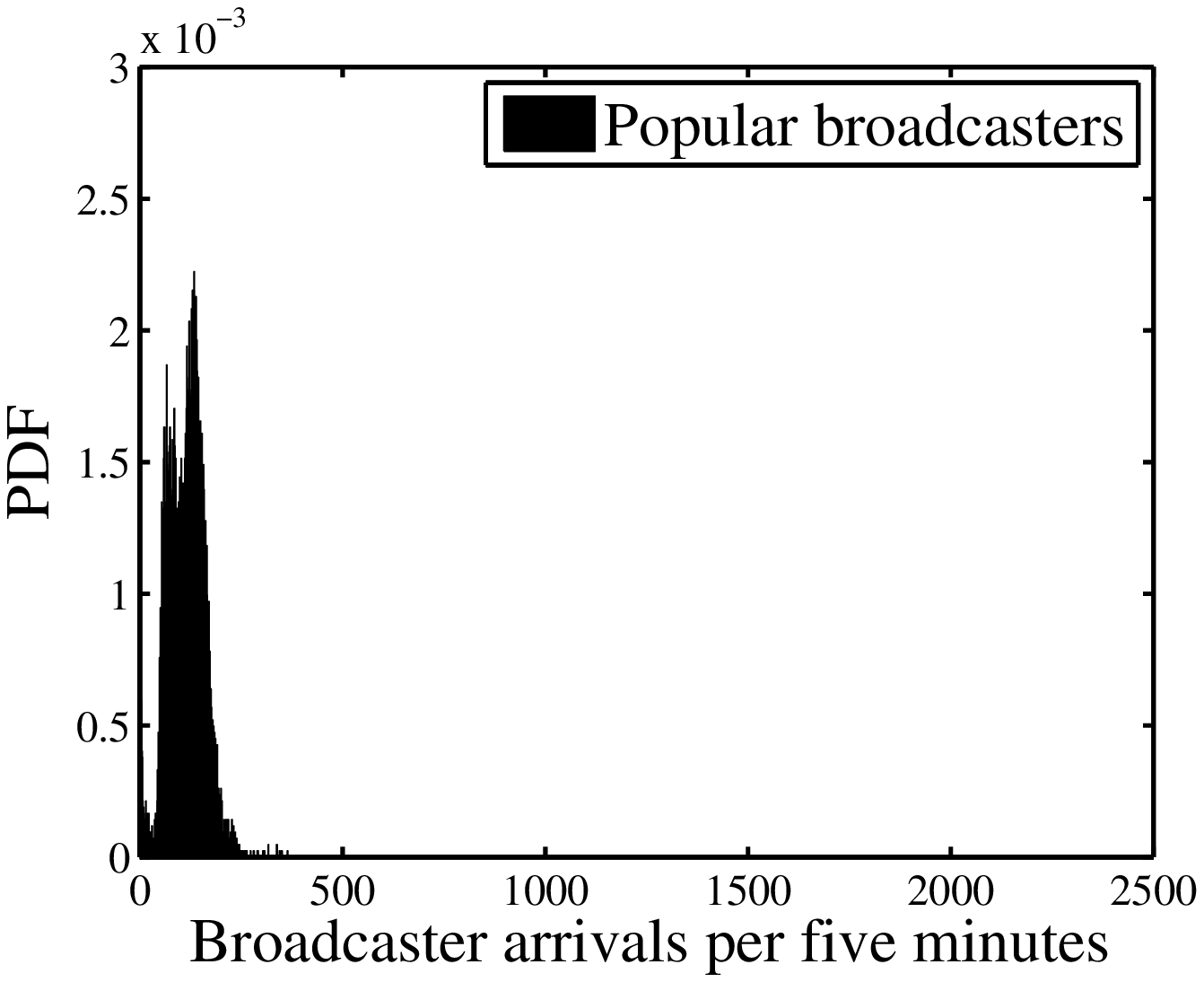}}
  \subfloat[Unpopular broadcasters]{
    \label{fig:arrivalb} %% label for second subfigure
    \includegraphics[width=0.48\textwidth]{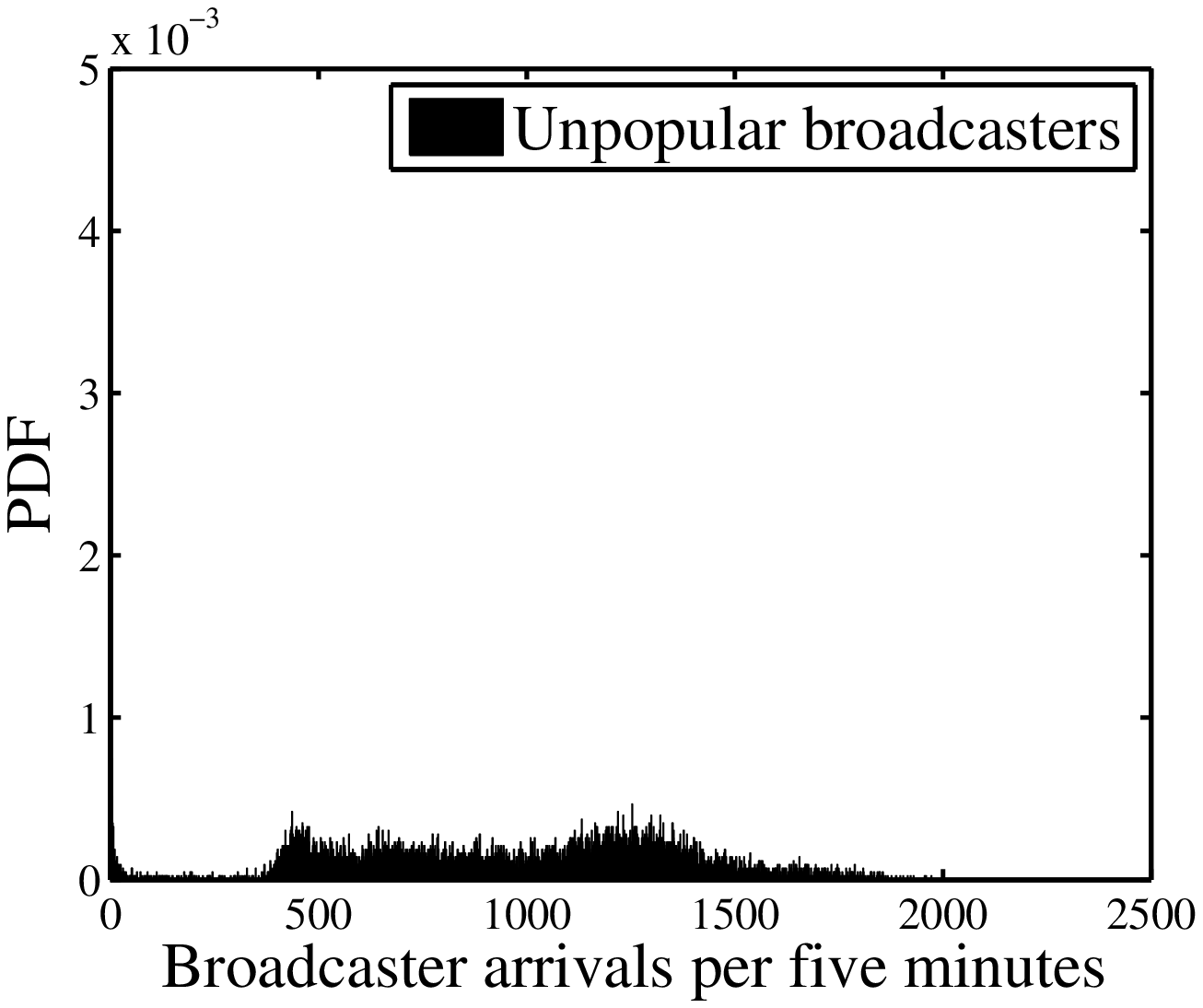}}
     \vspace{-2mm}
  \caption{Broadcaster arrivals per five minutes.}
  \label{fig:arrival} %% label for entire figure
\end{minipage}
\begin{minipage}[t]{0.5\linewidth}
 \centering
 \subfloat[Bandwidth consumption]{
    \label{fig:band} %% label for second subfigure
    \includegraphics[width=0.48\textwidth]{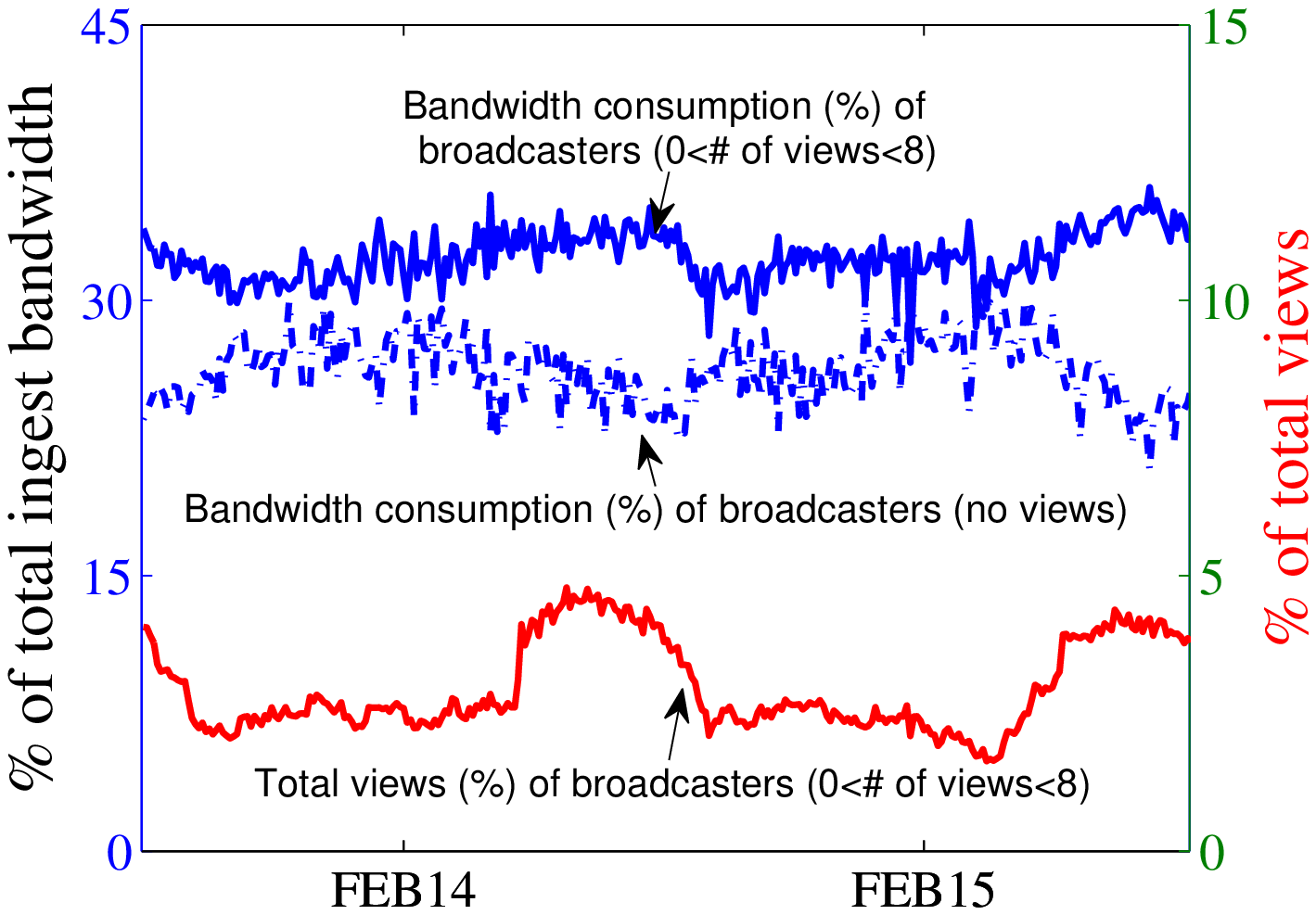}}
    \subfloat[Computation consumption]{
    \label{fig:cpu} %% label for first subfigure
    \includegraphics[width=0.48\textwidth]{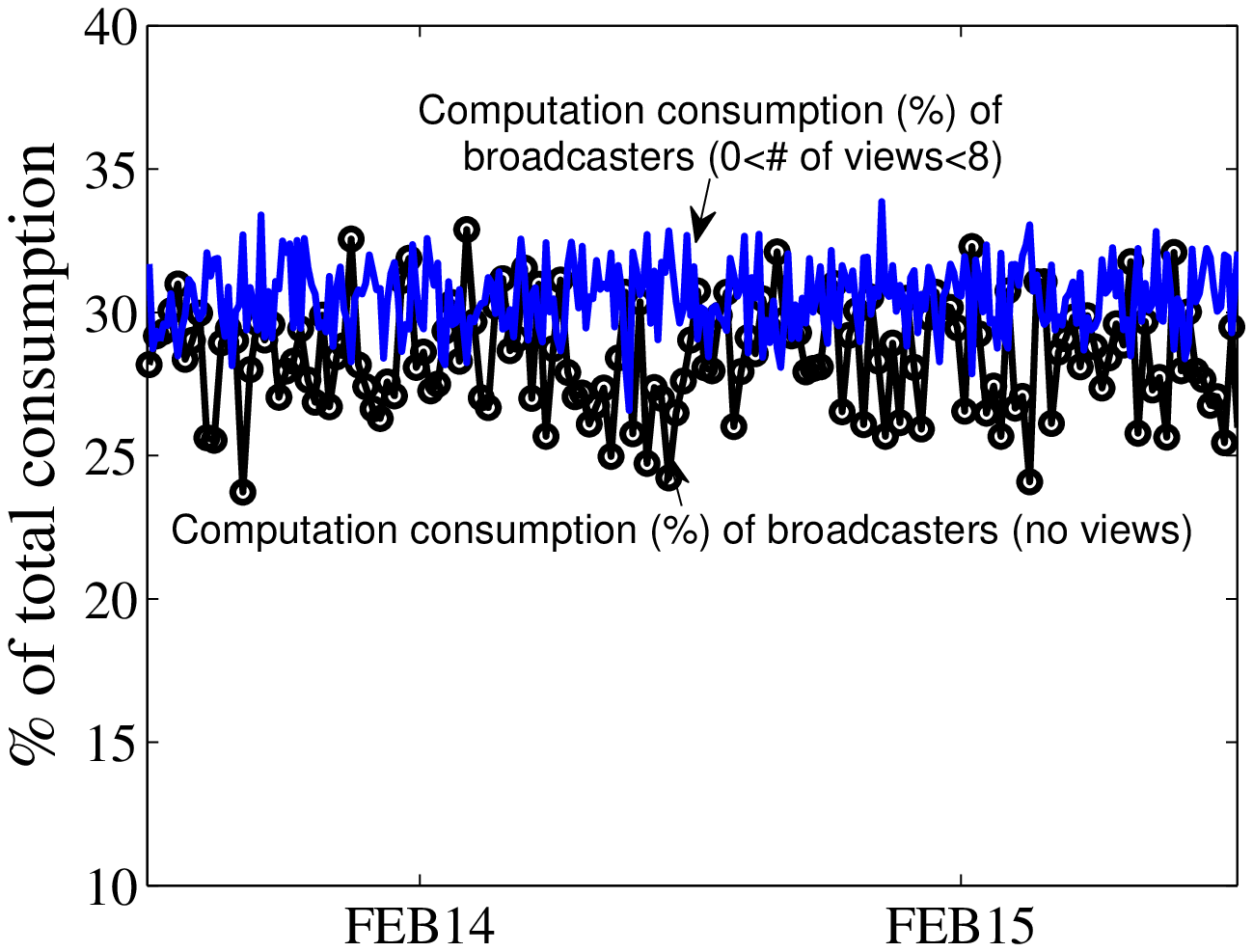}}
     \vspace{-2mm}
  \caption{The effectiveness of resource consumption.}
  \label{fig:consumption} %% label for entire figure
\end{minipage}
\vspace{-5mm}
\end{figure*}

\textit{Dynamics of Crowdsourced Live Broadcasters}: If the concurrent number of views in one stream is less than 8, the popularity of this stream is quite low. To further investigate the characteristics of them, we compare the distribution of their duration with popular streams in Figure~\ref{fig:plotduration}. This figure shows that the durations of 80\% of unpopular streams are less than 83 minutes, implying the workloads of them are highly dynamic. Because the number of unpopular streams is quite large (about 8.13 million), massive unpopular streams could occupy the datacenter resources frequently and dynamically. We also calculate the total duration of all unpopular streams in one month to be nearly 830 years, while the total duration of popular streams is only 310 years. Therefore, a huge amount of resources cannot be utilized effectively. To illustrate the different characteristics of two types of broadcasters, we plot their activities during ten days in Figure~\ref{fig:pop} and~\ref{fig:unpop}. Figure~\ref{fig:pop} shows that broadcaster A has a regular live schedule with the stable live duration, attracting a large number of viewers. While the broadcaster B in Figure~\ref{fig:unpop} not only has irregular schedules, but also consumes dedicated resources during the dynamic live duration. We also plot The Probability Distribution Function (PDF) of the broadcaster arrival rate every five minutes in Figure~\ref{fig:arrival}. This figure shows that the arrivals of popular broadcasters are clearly lower than 300, while the unpopular broadcaster's arrival rate has a considerable range from 400 to 1800. Due to the frequent arrivals and huge resource consumptions, it is necessary to enhance current CLS systems with optimizing the dynamic workloads of these unpopular broadcasters.

\textit{Challenge of Hosting Unpopular Broadcasters}: To evaluate the underlying challenges of these unpopular broadcasters, we use the playback bitrate and resolution of each live stream to estimate the consumptions of bandwidth and computation resources based on the measurement works in ~\cite{Aparicio-Pardo:2015:MMSys}. Figure~\ref{fig:consumption} shows the proportion of bandwidth/computation consumption of two types of broadcasters when they stream live content to ingesting servers on Feb 14th/15th, 2015. The broadcasters who do not have any viewers consume about 25\% (\textit{resp.} 28\%) of bandwidth (\textit{resp.} computation) resources. At the meantime, about 33\% (\textit{resp.} 31\%) of bandwidth (\textit{resp.} computation) resources are consumed by the broadcasters who only have less than $8$ concurrent viewers. This figure also shows that these broadcasters only attract less than 5\% concurrent viewers, which means that CLS service providers have to carry out a large number of ingesting servers to allocate these unpopular broadcasters dedicated bandwidth/computation resources continually.
\section{HyCLS Architecture}
\label{sec:design}
Based on our Twitch measurement, we have demonstrated that the characteristics of broadcasters/streams and illustrated that the dedicated resources are not to be consumed effectively. Our previous EC2-based measurements also illustrate public cloud can support crowdsourced live streaming effectively~\cite{report}. As such, we present the architecture of our hybrid cloud-assisted crowdsourced live streaming system HyCLS in this section.

In CLS systems, broadcasters constantly utilize the streaming pipeline, considering the latency-sensitive feature, any service interruption will generate a series of degradation of viewer's QoE. Therefore, the main challenge of our design is to optimize the broadcast latency. Besides, crowdsourced live events, wherein several broadcasters simultaneously start live-broadcast, have a more stringent requirement on the disparity of broadcast latencies between various broadcasters. To address these problems, our design focuses on the stream pipeline and optimizes the following three steps: (1)~\textit{Initial Offloading}, for the broadcasters who have historical information of live streams, the system makes an offloading decision between public clouds and dedicated datacenters when they start to connect ingesting servers. (2)~\textit{Ingesting Redirection}, according to the broadcasters' performances, the system assigns one alternative ingesting area and redirects her/his source streaming; (3)~\textit{Transcoding Schedule}, each offloading also considers the transcoding capacities of various service areas in HTTP Live Streaming scenario. In fact, step 2 and 3 have to be designed together, the reason is that once the workload of a broadcaster is offloaded to a certain ingesting area, transcoding workload has to be processed in the same area.

\begin{figure}[!ht]
 \vspace{-3mm}
\begin{minipage}[t]{0.55\linewidth}
 \centering
 \includegraphics[width=1\textwidth]{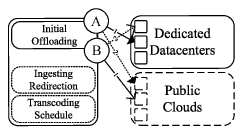}
 \vspace{-5.5mm}
  \caption{The design of HyCLS.}
  \label{fig:fw} %% label for entire figure
\end{minipage}
\begin{minipage}[t]{0.44\linewidth}
 \centering
 \includegraphics[width=0.9\textwidth]{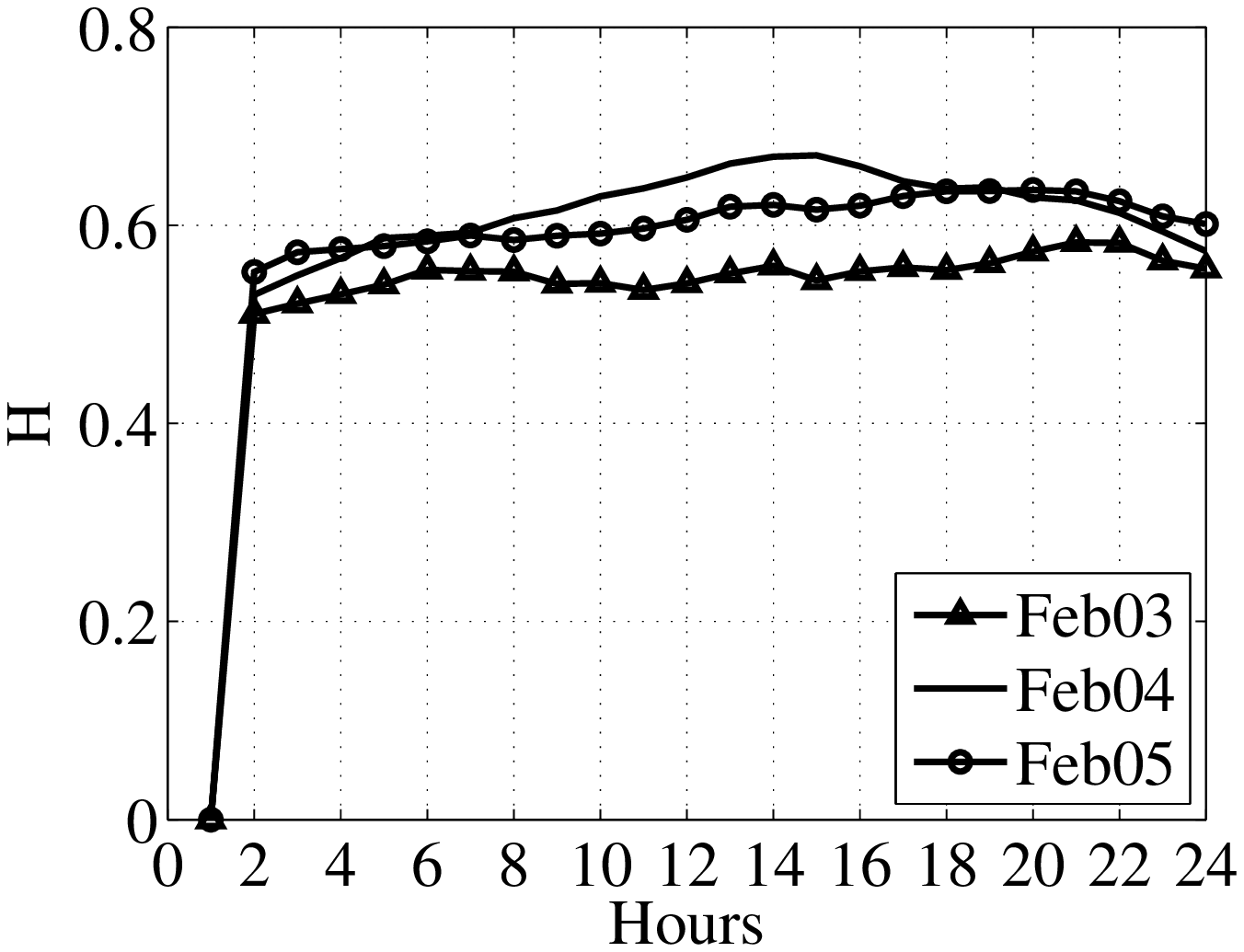}
 \vspace{-2mm}
  \caption{Impacts of $H$.}
  \label{fig:h} %% label for entire figure
\end{minipage}
\vspace{-3mm}
\end{figure}

We introduce our design as shown in Figure~\ref{fig:fw}, which represents the main components in the HyCLS system. For example, according to the historical information, Initial Offloading strategy first assigns broadcaster A and B to the second dedicated datacenter and the second public cloud area, respectively. After several time slots, Ingesting Redirection and Transcoding schedule modules migrate them to a proper service area based on our strategies. Next, we propose the design of offloading decision and optimize the Ingest Redirection and Transcoding Schedule together in Section~\ref{sec:formulation}.
In our hybrid framework, the first challenge is how to select proper ingesting server to broadcasters at the beginning of live-broadcast. Therefore, we have to estimate the stability of each broadcaster based on her/his historical activities. For one broadcaster $b$ who has activities in recent $n$ days ($n \geq 2$), we first divide the $i$th day to $m$ equal time slots, each time slot $j$ has a value $d_{i, j}$ that indicates whether $b$ have a live streaming in current time slot. In fact, $SI^{(b)}$ reflects the similarity of $b$'s resource consumption in recent $n$ days.

\begin{myequation}
\label{eq:u1}
SI^{(b)} =
  \begin{cases}
     \frac{1}{n} \sum_{i=2}^{n} \frac{\sum_{j=1}^{m} d_{i,j}^{(b)} \cdot d_{i-1,j}^{(b)}}{\sum_{j=1}^{m} d_{i-1,j}^{(b)}} &  \text{if $\sum_{j=1}^{m} d_{i-1,j}^{(b)} \neq 0$}\\
    0 &  \text{otherwise}
  \end{cases}
\end{myequation}

Given the stable index $SI^{(b)}$ of a broadcaster $b$, a straightforward way to give the offloading decision is to set a threshold $H$: if $SI^{(b)} \geq H$, $b$ will be assigned to the ingesting servers in dedicated datacenters, otherwise, public cloud ingests the live streaming of $b$. Using a firm threshold, however, suffers from the following drawback: during lower workload stage, leasing public clouds to specifically ingest unpopular workloads is not a cost-effective strategy, because the existing resources in dedicated datacenters can completely process every broadcaster's live stream. We solve this problem by detecting the existing broadcasters' $SI$ in one dedicated datacenter and update the value of $H$ to the average $SI$ of all broadcasters per unit time. Followed by the growth of broadcasters, more and more regular broadcasters will be ingested into dedicated datacenters, and other dynamic broadcasters are offloaded to public clouds at the beginning of live-broadcast. We further evaluate the effectiveness of the stable index in Section~\ref{sec:eva}.

\section{Problem Formulation and Solution}
\label{sec:formulation}
Due to the dynamic and unpredictable features of broadcasters in CLS systems, we design the ingesting and transcoding strategies based on the current workloads status. The workloads can be migrated among  different service areas in real-time. Considering the critical features including crowdsourced live events and latency synchronization, we take broadcast latency as our objective and propose a formal description of our optimization problem in the current crowdsourced scenario.
\subsection{Problem Formulation}

To make the problem easy to discuss, we quantize time into discrete time slots, which may be a few minutes to several hours (e.g., five minutes in our experiment). We use $B^{(t)}$ to denote the set of broadcasters and $E^{(t)}$ to denote the set of crowdsourced live events in time slot $t$. ($\forall i = 1,2,\cdots,m, \forall j = 1,2,\cdots,m$, $e_{i} \in E^{(t)}$, $|e_{i}| \geq 1$, $e_{i}\cap_{i \neq j} e_{j} = \varnothing$, and $\cup e_{i} = B^{(t)}$). We define $R$ as the set of ingesting areas where a broadcaster can be connected to upload live content and define set $W^{(t)}_{r}$ as the bandwidth demand of ingesting area $r$. We assume that the instance in public cloud are homogeneous and let $ \mathcal{W}$ denote the bandwidth capacity of each instance. Therefore, we do not consider the optimization inside each service area, a large number of works have focused on this area and acquired a better optimization~\cite{Feng:2012:INFOCOM}.

We consider the impacts of ingesting stage and transcoding workload with various versions on different ingesting areas. As such, $b$'s broadcast latency $L_{(b,r,v)}^{(t)}$ is calculated as:
\begin{myequation}\label{eq:brv}
L_{(b,r,v)}^{(t)} = l_{(b,r)}^{(t)}+ l^{(t)}_{(q_{b},q_{v})} + l_{(r,v)}
\end{myequation}
\noindent where $v$ is the transcoding version, $v \in V$, $l_{(b,r)}^{(t)}$ is the link latency between $b$ and $r$, $l_{r}^{(t)}$ is the ingesting latency that is determined by the instance type in $r$, $q_{b}$ and $q_{v}$ are bitrates of source (i.e., broadcaster $b$) and transcoding version $v$ ($v \in V, V = {0, \mathbb{Z}^{+}}$), respectively. $l^{(t)}_{(q_{b},q_{v})}$ is the transcoding latency, which can be measured in advance. $l_{(r,v)}$ is the latency between ingesting area $r$ to a class of viewers $v$, which will be defined in next section.

We now define utility function $U^{(t)}(b,r,v)$ as:
\begin{myequation}\label{eq:su}
U^{(t)}(b,r) = \sum_{v \in V} G^{(t)}(b,r,v) \cdot N^{(t)}_{(b,v)}
\end{myequation}
\noindent where $N^{(t)}_{(b,v)}$ is the number of viewers who watch $b$'s $v$ version streaming in this time slot. This value is initially determined by $b$'s historical distribution of different versions. $G^{(t)}(b,r,v)$ means the gain when $b$ selects $r$ as the ingesting and transcoding area and is calculated as follows:
\begin{myequation}\label{eq:second}
\begin{aligned}
G^{(t)}(b,r,v) &= \alpha + ln(1-\beta L_{(b,r,v)}^{(t)}) \\
&= \alpha + ln(1-\beta ( l_{(b,r)}^{(t)} + l^{(t)}_{(q_{b},q_{v})} + l^{(t)}_{(r,v)}) )
\end{aligned}
\end{myequation}
\noindent where $l^{(t)}_{(q_{b},q_{v})}$ denotes the transcoding latency. If $q_{b} \leq q_{v}$, $l^{(t)}_{(q_{b},q_{v})} = l^{(t)}_{(q_{b},q_{b})}$, which depends on the current computing capacity of area $r$ and is monotonously increasing on both $q_{b}$ and $q_{v}$~\cite{Wu:2013:SIGCOMM}.

Based on previous definitions, our objective is :
\begin{myequation}\label{eq:newobj}
\underset{e \in E^{(t)}}{\text{Maximize  }} F(A^{(t)}) = \min_{\substack{b \in e \\ r \in R}} \{  U^{(t)}(b,r)\}
\end{myequation}
\textit{subject to:}
Resource Availability Constraints:
\begin{myequation}\label{eq:bac}
\forall r \in R,   W^{(t)}_{r} \leq \mathbb{W}_{r}
\end{myequation}
\begin{myequation}\label{eq:cac}
\forall r \in R,   C^{(t)}_{r} \leq \mathbb{C}_{r}
\end{myequation}
Budget Constraints:
\begin{myequation}\label{eq:bbc}
\sum_{r \in R} \frac{ W^{(t)}_{r}}{\mathcal{W}}\cdot Cost_{w}(r) \cdot I(r)  \leq K_{w}
\end{myequation}
\begin{myequation}\label{eq:cbc}
\sum_{r \in R} \frac{ C^{(t)}_{r}}{\mathcal{C}} \cdot Cost_{c}(r)   \cdot I(r) \leq K_{c}
\end{myequation}

\noindent where $\mathbb{W}_{r}$ is the bandwidth capacity of ingesting area $r$. $Cost_{w}(r)$ is the bandwidth price in area $r_{i}$. The bandwidth constraint~\eqref{eq:bac} asks that at any given time, the bandwidth demands have to be satisfied. $C^{(t)}_{r}$ is the computing demand of area $r$, $\mathcal{C}$ denotes the amount of per unit computing resource. $Cost_{c}(r)$ is the instance price in $r$ in terms of computing capacity. The computing constraint~\eqref{eq:cac} guarantees that at any given time $t$, the computation resource consumption of each transcoding task can be satisfied. The budget constraints~\eqref{eq:bbc} and ~\eqref{eq:cbc} guarantees that the bandwidth/computation cost is lower than the budget $K_{c}$/$K_{w}$, which we assume can at least serve all offloading workloads.

\subsection{Solution}

Current formulated objective~\eqref{eq:newobj} has four constraints~\eqref{eq:bac}~\eqref{eq:cac}~\eqref{eq:bbc}, and ~\eqref{eq:cbc}. It is hard to solve this optimization problem efficiently in a short time. Fortunately, the bandwidth cost and computation cost are not independent due to the pricing criteria of the instance on a public cloud. Previous studies on EC2 instances already reveal that the bandwidth capacity is more than 700Mbps on m3.large instance~\cite{m3m}. As such, based on our measurement results in~\cite{report}, generating low-latency live streams will consume a vast of computation resources. If we relax constraints~\eqref{eq:bac} and~\eqref{eq:bbc}, another constraints still work for optimizing objective~\eqref{eq:newobj}. Assuming that the capacities of the different public cloud service area are given, our assignment problem can therefore be transformed into a 0-1 Multiple Knapsack problem, which is known to be NP-hard~\cite{knapsack}. Although the optimal solution can be reached through meticulously searching all possible assignments, this is unpractical in real CLS systems. Inspired by our previous work~\cite{Wang:2014:ICNP}, we thus propose a heuristic algorithm, which consists of scaling decrease and resource assignment, as shown in Algorithm~\ref{alg:1}. In the scaling decrease step (line 1 to 10), we eliminate the redundant assignment solutions based on the optimization target. The line~\ref{line:1-a} search the maximum value from the set of a minimum utility of each assignment $(b,r,v)$ in crowdsourced live events. The line~\ref{line:1-b} to~\ref{line:1-c} then remove all useless assignment and guarantee the rest of assignment can be effectively used in the next algorithm. We next use resource assignment (line 11 to 22) to implement an effective solution. The main idea is to utilize the utility in live streaming broadcast latency by a unit of computation resources.

\begin{algorithm}[htb]
\scriptsize
\caption{WorkloadAssignment()}
\label{alg:1}
\begin{algorithmic}[1]
\FOR {each crowdsourced live event $e \in E$}
    \STATE $U^{(t)}_{e} \leftarrow \max_{r \in R} \{ \min_{\substack{b \in e }} \{  U^{(t)}(b,r)\}\}$ ;
    \label{line:1-a}
\ENDFOR
\FOR {each crowdsourced live event $e \in E$}
    \label{line:1-b}
    \FOR {each assignment $(b,r) \in A^{(t)}(b,r)$}
        \IF {$U^{(t)}(b,r) < U^{(t)}_{e}$ and $IsPath(b,r) == true$}
            \STATE $A^{(t)}_{*} \leftarrow  A^{(t)} - (b,r)$;
                //Remove this assignment path
        \ENDIF
    \ENDFOR
\ENDFOR
\label{line:1-c}
\STATE Sort $(b,r)$ by descendant order of $U^{(t)}(b,r)/Cost_{c}(r)$;
\FOR {each assignment $(b,r) \in A^{(t)}_{*}$ }
    \STATE $r_{sorted} \leftarrow$ Sorted available area $r$ of $b$ by descendant order of $U^{(t)}(b,r)$;
    \FOR {each $r \in r_{sorted}$ }
        \IF {$C_{r}^{(t)} - c(b) \geq 0 $}
            \STATE $A^{(t)}_{*} \leftarrow  A^{(t)}_{*} - {(b,\cdot)}$;
                //Remove all assignment of $b$
            \STATE $C_{r}^{(t)} \leftarrow C_{r}^{(t)} + c(b)$;
                //$c(b)$ is the computation consumption of transcoding workloads $b$
            \STATE $A^{(t)}_{*} \leftarrow  A^{(t)}_{*} + (b,r)$;
        \ENDIF
    \ENDFOR
\ENDFOR
\RETURN $A^{(t)}_{*}$
\end{algorithmic}
\end{algorithm}

\section{Performance Evaluation}
\label{sec:eva}
We now evaluate the performance of our solution via trace-based simulation, which captures the broadcasters' streaming pattern, including resolution, partner status, and concurrent viewers, etc., in Twitch dataset. We consider two broadcaster's types: partner, whose live streaming can be adaptively transcoded, and common broadcasters, whose viewers only can watch the source quality HTTP Live Streaming. At the meantime, we make a few simplifications in the simulation based on realistic settings: first, to simplify the complexity of algorithm, we consider that the EC2 instances are homogeneous (m3.large) and latency $l_{(r,v)}$ is fixed for a certain quality level of HTTP Live Streaming; second, due to the confidential nature of official implementation, we cannot acquire the details of dedicated datacenter, we show the comparisons of extra outlay when workloads are offloaded into public cloud. The price data of instances come from Amazon. The following settings are the default parameters in the simulation: to normalized the impacts of broadcast latency, we set $\alpha = 1$ and $\beta = 0.011$, which makes the gain $G^{(t)}(\cdot) \in [0,1]$, if the broadcast latency $L^{(t)}_{(\cdot)} \in [0, 57]$, which embraces a general broadcast latency interval $[10, 40]$ in Twitch~\cite{Zhang:2015:NOSSDAV}. The algorithms are launched per five minutes, which also is the time slot of crawling data.

We first conduct simulations to study the impacts of stable index $SI$ and threshold $H$. We set $n = 2$ to calculate the broadcaster's stable index in advance and set the initial threshold $H = 0$. To illustrate the efficiency of this threshold, we use it to classify the new broadcasters without any other strategies. We assume that the offloading starts when the bandwidth consumption is up to 60\% of the dedicated datacenter. Figure~\ref{fig:h} illustrates the evolution of $H$ and its impacts for the public cloud during three days (Feb 3rd-5th, 2015). From this figure, we observe that the value of $H$ increase dramatically at the beginning of that day, and then it stables between $0.5$ and $0.7$. At the peak traffic time (from 9:00AM to 13:00PM), a vast of broadcasters arrive streaming systems; therefore, the value of $H$ occurs a small decrease. The limitation of $H$, however, induces that public cloud only hosts a few number (maximum 6.5\%) of broadcasters. Thus, threshold $H$ plays a beneficial role in the offloading process, but it still cannot reduce the impacts of dynamic broadcasters sufficiently.

\begin{figure}[!ht]

\begin{minipage}[t]{1\linewidth}
 \centering
  \subfloat[Lease cost of three approaches]{
    \label{fig:eva2a} %% label for first subfigure
    \includegraphics[width=0.48\textwidth]{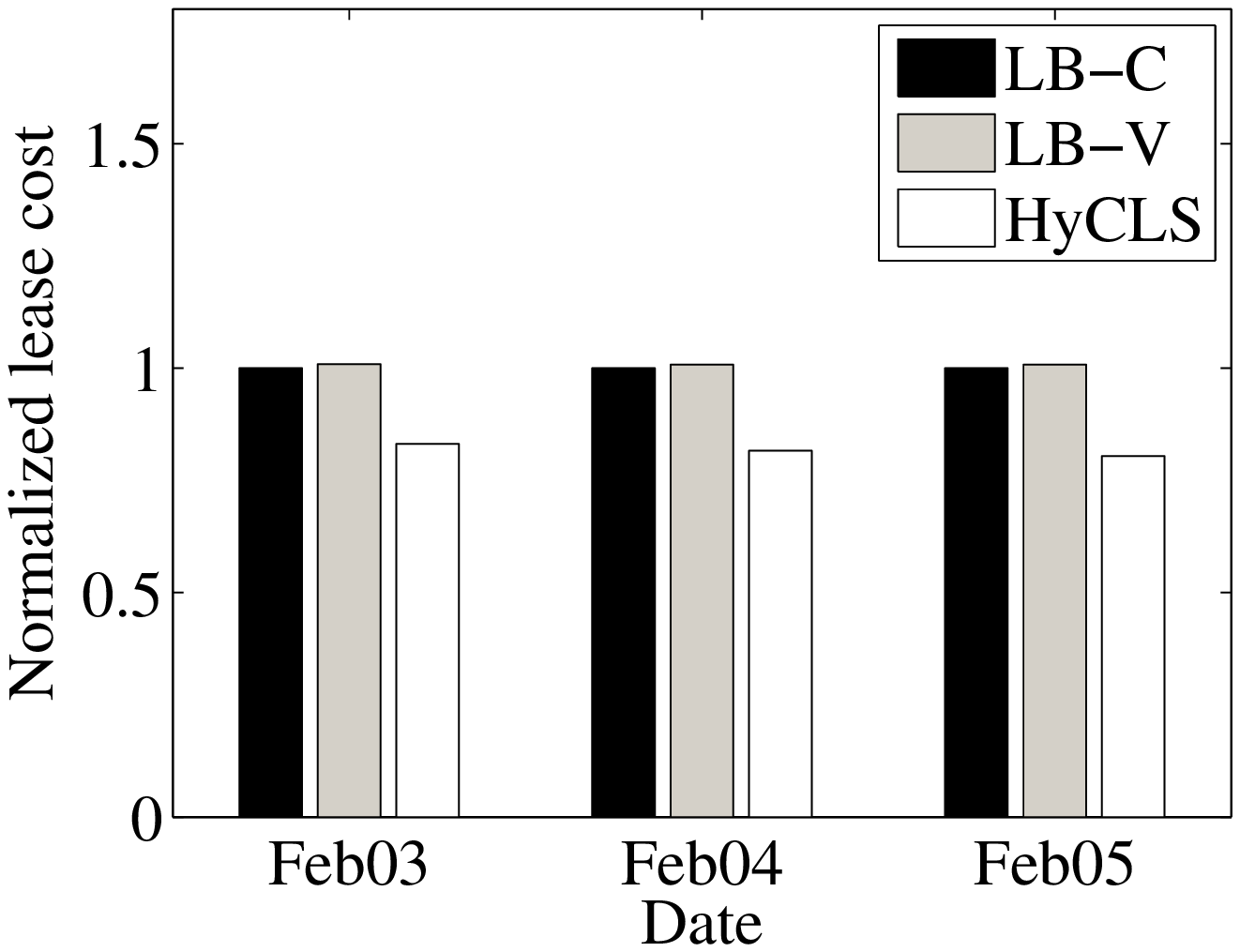}}
  \subfloat[Migration performance (Feb.03-28)]{
    \label{fig:eva2b} %% label for second subfigure
    \includegraphics[width=0.48\textwidth]{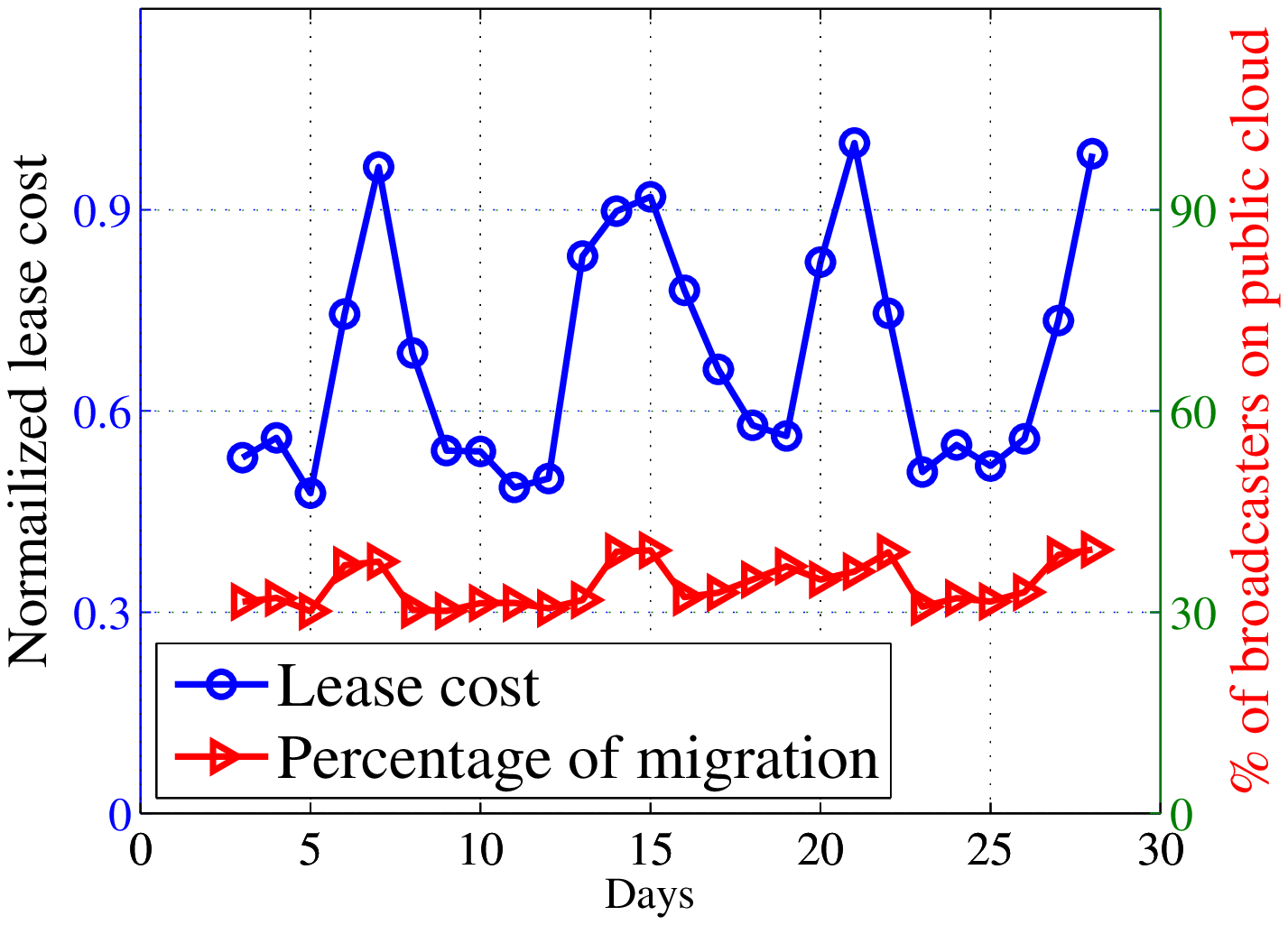}}

  \caption{The performance of proposed solutions.}
  \label{fig:eva2} %% label for entire figure
\end{minipage}
\end{figure}

With the previous parameter setting of $H$, we then conduct simulations to investigate how HyCLS performs with the real data traces.
Figure~\ref{fig:eva2a} compares the lease cost of three workload provisioning approaches: views-based (LB-V), computation-based (LB-C), and HyCLS-based approaches in three days. LB-V only considers the current number of views in different live streams, while LB-C migrates workload based on the consumption of computation resources. For ease of comparison, the lease cost in each day is normalized by the corresponding cost of the LB-C approach. Our HyCLS-based approach has the lowest cost, decreasing 16.9\%-19.5\% of LB-C approach and 17.8\%-20.4\% of LB-V approach. Another observation is that the lease cost of Feb03 is higher than those of another two days in all approaches. This is because there has the highest number of broadcasters on Feb03. We also plot the normalized lease expenses and the average percentage of migration during our whole datasets in Figure~\ref{fig:eva2b}, we can observe that the daily decreasing cost performs the weekly pattern and provide elastic workload provisioning cost-effectively. Moreover, more than 30\% of broadcasters are migrated to the public cloud in every day. We further explore the highest percentage of migration and find that up to 59.9\% of broadcasters will be assigned to public clouds on several time slots. Our simulation results show that compared with spending massive outlays to manage and upgrade dedicated datacenters, leasing flexible public cloud is a cost-effective solution in terms of decreasing the influences of dynamic unpopular broadcasters and providing highly available live streaming services.

\section{Conclusion and Future Work}
\label{sec:conclusion}
This paper presented HyCLS, a generic framework that facilitates migrating crowdsourced live streaming between dedicated datacenters and public clouds. We strived to offer the comprehensive understandings on the practical crowdsourced live streaming system and explore the potential enhancement. We first measured Twitch-based datasets to investigate the challenges therein. We observed that unpopular broadcasters consume the massive valuable dedicated resources continually. We then proposed a hybrid design for the initial offloading, as well as dynamic ingesting redirection and transcoding assignment that accommodates unpredictable workloads and realizes the adaptive offloading in demand. Extensive simulations driven by traces from Twitch and settings from Amazon EC2 demonstrated the cost-effectiveness and superior migration of HyCLS. We are currently examining the performance of our hybrid cloud design on the PlanetLab. We are also interesting in developing a better initial offloading strategy, as well as incorporating other factors into stable index and threshold evolution, e.g., broadcaster's social characteristics.

\section*{Acknowledgment}
This research is supported by an NSERC Discovery Grant, NSERC RTI Grant and NSERC Strategic Grant.
C. Zhang' s work is supported in part by a China Scholarship Council (CSC) Scholarship Program.
H. Wang's work was supported by Chancellor's Small Grant and Grant-in-aid programs from the University of Minnesota.

\bibliographystyle{abbrv}
\bibliography{nossdav_16_arxiv}

\begin{thebibliography}{10}

\bibitem{Aparicio-Pardo:2015:MMSys}
R.~Aparicio-Pardo, K.~Pires, A.~Blanc, and G.~Simon.
\newblock Transcoding live adaptive video streams at a massive scale in the
  cloud.
\newblock In {\em ACM MMSys, 2015}.

\bibitem{Carlier:2011:MM}
A.~Carlier, G.~Ravindra, V.~Charvillat, and W.~T. Ooi.
\newblock Combining content-based analysis and crowdsourcing to improve user
  interaction with zoomable video.
\newblock In {\em ACM MM, 2011}.

\bibitem{knapsack}
A.~Drexl.
\newblock A simulated annealing approach to the multiconstraint zero-one
  knapsack problem.
\newblock {\em Computing}, 40(1):1--8, 1988.

\bibitem{report:esports}
A.~Gaudiosi.
\newblock esports: Espn's analysis on esport trends.
\newblock http://goo.gl/p2aSXZ, July 2015.

\bibitem{m3m}
M.~Hajjat, R.~Liu, Y.~Chang, T.~E. Ng, and S.~Rao.
\newblock Application-specific configuration selection in the cloud: impact of
  provider policy and potential of systematic testing.
\newblock In {\em IEEE INFOCOM, 2015}.

\bibitem{Kaytoue:2012:WWW}
M.~Kaytoue, A.~Silva, L.~Cerf, W.~Meira, Jr., and C.~Ra\"{\i}ssi.
\newblock Watch me playing, i am a professional: A first study on video game
  live streaming.
\newblock In {\em ACM WWW, 2012}.

\bibitem{Liu:2009:p2p}
Z.~Liu, C.~Wu, B.~Li, and S.~Zhao.
\newblock Why are peers less stable in unpopular p2p streaming channels?
\newblock In {\em NETWORKING 2009}, volume 5550 of {\em Lecture Notes in
  Computer Science}, pages 274--286. 2009.

\bibitem{Motoyama:2010:USENIX}
M.~Motoyama, K.~Levchenko, C.~Kanich, D.~McCoy, G.~M. Voelker, and S.~Savage.
\newblock Re: Captchas: Understanding captcha-solving services in an economic
  context.
\newblock In {\em USENIX Security, 2010}.

\bibitem{Rawat:2014:MM}
Y.~S. Rawat and M.~S. Kankanhalli.
\newblock Context based photography learning using crowdsourced images and
  social media.
\newblock In {\em ACM MM, 2014}.

\bibitem{Feng:2012:INFOCOM}
F.~Wang, J.~Liu, and M.~Chen.
\newblock Calms: Cloud-assisted live media streaming for globalized demands
  with time/region diversities.
\newblock In {\em IEEE INFOCOM, 2012}.

\bibitem{Wang:2014:ICNP}
H.~Wang, R.~Shea, X.~Ma, F.~Wang, and J.~Liu.
\newblock On design and performance of cloud-based distributed interactive
  applications.
\newblock In {\em IEEE ICNP, 2014}.

\bibitem{Wikimapia}
Wikimapia.
\newblock Wikimapia - let's describe the whole world!
\newblock http://wikimapia.org/, July 2015.

\bibitem{Wu:2013:SIGCOMM}
Y.~Wu, C.~Wu, B.~Li, and F.~C. Lau.
\newblock vskyconf: Cloud-assisted multi-party mobile video conferencing.
\newblock In {\em ACM SIGCOMM Workshop on MCC, 2013}.

\bibitem{report}
C.~Zhang.
\newblock Performance analysis of {A}mazon {EC2} in {CLS} scenario.
\newblock https://goo.gl/c7CTOa, Feb 2016.

\bibitem{Zhang:2015:NOSSDAV}
C.~Zhang and J.~Liu.
\newblock On crowdsourced interactive live streaming: A twitch.tv-based
  measurement study.
\newblock In {\em ACM NOSSDAV, 2015}.

\end{thebibliography}

%\begin{figure}[!t]
%  \center
%  \includegraphics[width=0.4\textwidth]{figure3-networking-qoe.eps}
%  \caption{The impacts of the networking condition at the receiver-side}
%  \label{fig:networking}
%\end{figure}

\end{document}